\documentclass[aps,prl,reprint,groupedaddress]{revtex4-1}

\usepackage{graphicx}
\usepackage{amsmath}
\begin{document}

\title{Direct measurement of sub-10 fs relativistic electron beams with ultralow emittance}

\author{Jared Maxson}
\email[]{jmaxson@ucla.edu}
\author{David Cesar}
\author{Giacomo Calmasini}
\author{Alexander Ody}
\author{Pietro Musumeci}
\affiliation{Department of Physics and Astronomy, UCLA, Los Angeles, California 90095, USA}

\author{David Alesini}
\affiliation{INFN-Laboratori Nazionali di Frascati, Via Enrico Fermi 40, 00044 Frascati, Rome, Italy}

\date{\today}

\begin{abstract}
Ultralow emittance ($\leq20$ nm, normalized) electron beams with $10^5$ electrons per bunch are obtained by tightly focusing an ultrafast ($\sim$ 100 fs) laser pulse on the cathode of a 1.6 cell radiofrequency photoinjector. Taking advantage of the small initial longitudinal emittance, a downstream velocity bunching cavity is used to compress the beam to $<10$ fs rms bunch length. The measurement is performed using a thick high voltage deflecting cavity which is shown to be well-suited to measure ultrashort durations of bunching beams, provided that the beam reaches a ballistic longitudinal focus at the cavity center.
\end{abstract}


\maketitle 

Crossing the 10 fs threshold in electron bunch length can enable breakthrough opportunities for compact electron sources, with applications ranging from ultrafast electron diffraction (UED) and microscopy (UEM), to advanced acceleration schemes. In UED/M, ultrashort, ultrabright beams are required to probe atomic motion at fundamental time scales. A wide array of photoemission accelerator technology has been developed or adapted for UED/M, ranging from keV \cite{ OudheusdenJongGeerEtAl2007, OudheusdenPasmansGeerEtAl2010,  ChatelainMorrisonKlarenaarEtAl2014, Ihee2001, Siwick2003, GliserinWalbranKrauszEtAl2015} up to MeV scale \cite{CesarMaxsonMusumeciEtAl2016, MusumeciMoodyScobyEtAl2010, WeathersbyBrownCenturionEtAl2015, ZhuZhuHidakaEtAl2015, LiTangDuEtAl2009, FilippettoQian2016, Manz2015}, which have pushed the temporal resolution of such instruments below 100 fs.  Shorter electron beams (sub-10 fs) are ultimately required to probe the fastest dynamics in solid state systems and directly observe bond-breaking in gas phase molecular reactions \cite{IshiokaHaseKitajimaEtAl2008, WeathersbyBrownCenturionEtAl2015, GliserinWalbranKrauszEtAl2015}.

Similarly, low emittance, ultrashort electron beams are critical for the development of plasma \cite{EsareySchroederLeemans2009, ChenDawsonHuffEtAl1985} and direct laser-based advanced acceleration schemes \cite{PeraltaSoongEnglandEtAl2013,WoottonWuCowanEtAl2016}. This is because such schemes utilize few fs longitudinal and micron scale transverse acceleration apertures, defined by either the plasma or laser wavelength, in order to provide GV/m gradients. Thus, injection of an external beam into these accelerators requires commensurately short bunch lengths and small 6D emittances \cite{DordaAssmannBrinkmannEtAl2016}.

An attractive method to achieve short bunch lengths in compact beamlines is velocity bunching (sometimes also termed ballistic bunching), which allows compression of an electron beam by a large factor with respect to the laser pulse duration used in the photoemission process. In a velocity bunching scheme, a negative longitudinal position-velocity correlation is established within a particle bunch (typically via an rf cavity), leading to a longitudinal focus downstream \cite{AndersonMusumeciRosenzweigEtAl2005, FerrarioAlesiniBacciEtAl2010}. Velocity bunching is most effective for low energy electrons (keV) \cite{OudheusdenJongGeerEtAl2007}, but can be performed on meter scales with electron beams of a few MeV. Nonlinear longitudinal phase space (LPS) correlations and space charge repulsion typically limit the shortest bunch lengths achievable. Both non-relativistic \cite{OudheusdenPasmansGeerEtAl2010, GliserinWalbranKrauszEtAl2015} and relativistic \cite{LuTangLiEtAl2015} velocity bunching experiments have to date demonstrated well below 100 fs rms bunch lengths, but have not yet yielded measurements of bunch lengths in the single digit fs. This is also due to challenges associated with beam diagnostics for very short bunch durations \cite{DornmairSchroederFloettmannEtAl2016}.

In this work, we describe a compact beamline setup by which we generate beams of unprecedented bunch length ($<$10 fs rms) and simultaneously very low transverse normalized emittance ($\leq 20$ nm), utilizing an rf gun and bunching cavity at the UCLA Pegasus laboratory. Exploring the tradeoffs between transverse and longitudinal initial emittances with particle tracking simulations, we find that the use of low charge (10$^5$ electrons), but very high phase space density beams enables us to break the 10 fs bunch length barrier. An additional important result of this paper is the demonstration that thick deflecting cavities are well suited for the direct measurement of these beams due to a transverse kick cancellation which occurs when the beam undergoes a non-laminar longitudinal focus (where the bunch head becomes the tail, and vice versa) inside the cavity.

\begin{figure*}
\includegraphics[width=0.8\textwidth]{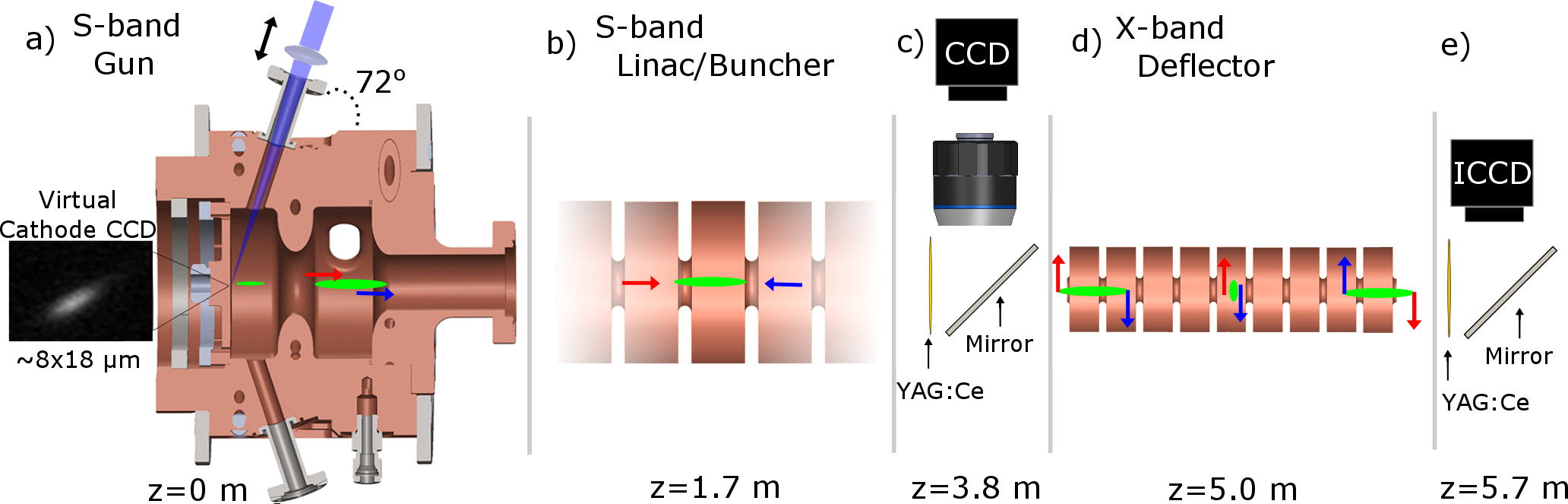}
\caption{Experimental layout with element positions (not to scale). a) The UV drive laser reaches a final focus at the copper photocathode of a 1.6 cell S-band photogun. The inset shows an image of the laser intensity distribution at the photocathode plane. b) An 11 cell linac is used as a buncher by applying a negative z--$\gamma$ correlation to the bunch. The red and blue arrows signify the applied force on the tail and head of the electron bunch, respectively. c) An in-vacuum microscope objective images a 20 $\mu$m thick YAG:Ce screen onto a CCD in air, providing a high spatial resolution (1.4 $\mu$m/pixel) beam profile monitor for emittance measurements. d) The electron bunch reaches a nonlaminar longitudinal focus at the center of an 9 cell X-band deflection cavity. e) The streaked electron profile is detected on a 50 $\mu$m thick YAG:Ce with an intensified camera. \label{gunfig} Two focusing solenoids (not shown) are located at $z$ = 0.24 m and 3.3 m.}
\end{figure*}

High frequency electron sources, such as the 1.6 cell S-band (2.856 GHz) photogun employed here \cite{AlesiniBattistiFerrarioEtAl2015}, achieve higher photocathode extraction fields than low frequency guns (such as dc sources or MHz class rf guns) and hence have the capability to utilize a smaller source size for a given charge. The resulting smaller emittance does not just benefit the beam transverse quality, it also enables the production of shorter bunch lengths by reducing the time-of-flight differences of radially separated particles in bunching schemes \cite{LoosGeerSavelievEtAl2006}, along with alleviating bunch length measurement-corrupting transverse effects in deflecting cavities \cite{FloettmannParamonov2014}, as discussed below. 

To generate nm-scale normalized emittances, we focus the drive laser (266 nm) onto a copper photocathode by utilizing a $72^\circ$ oblique incidence vacuum port and a final focus lens ($f = 175$ mm) mounted on a translation stage. This is depicted in Fig. 1(a). A beam splitter (not shown) is used to monitor the laser spot size on a fluorescent screen (YAG:Ce) located at the virtual photocathode plane. The use of oblique incidence minimizes the lens-photocathode distance and allows us to obtain a $8 \times 18$ $\mu$m intensity root mean square (rms) cathode spot size as shown in Fig. 1a. Taking into account the optical point spread function (PSF) of the virtual cathode screen, this is considered an upper bound on the actual photocathode spot size. For a copper photocathode with an intrinsic emittance of 0.8 $\mu$m/mm at 266 nm \cite{LiRobertsScobyEtAl2012}, this corresponds to an initial emittance upper bound of $\sqrt{\epsilon_{nx} \epsilon_{ny}} = 10$ nm. For a $E_0 = $50 MV/m extraction field and a short laser duration on the cathode producing a ``pancake" aspect ratio bunch, this spot size sets the space charge limited bunch charge to $ 2 \pi E_0 \epsilon_0 \sigma_x \sigma_y =$ 400 fC, suggesting optimal operation in the 10s of fC \cite{MusumeciMoodyEnglandEtAl2008}.

To demonstrate the ultralow transverse emittances possible with this laser geometry, we first used a relatively long $\sigma_{t, uv} \approx$ 1.1 ps rms temporal laser width, which provides a bunch length after emission $\frac{1}{2 m} e E_0\sigma_{t, uv}^2 \approx 5 \mu$m, comparable to the transverse size, a mode which has been shown \cite{FilippettoMusumeciZolotorevEtAl2014, LiRobertsScobyEtAl2012} to alleviate space charge forces and emittance growth in emission and transport. Upon reaching a transverse waist from the solenoid focusing lens near the gun, the bunch energy is boosted to 8 MeV with an 11 cell linac \cite{BarovChangMillerEtAl2012}, which will later also serve as the bunching cavity. The emittance is measured with a second solenoid and high spatial resolution (1.4 $\mu$m/pixel) profiler downstream of the linac. High spatial resolution is achieved utilizing a thin (20 $\mu$m) YAG:Ce crystal with an in-vacuum infinity corrected microscope objective coupled to an in-air CCD.

\begin{figure}
\includegraphics[width=0.4\textwidth]{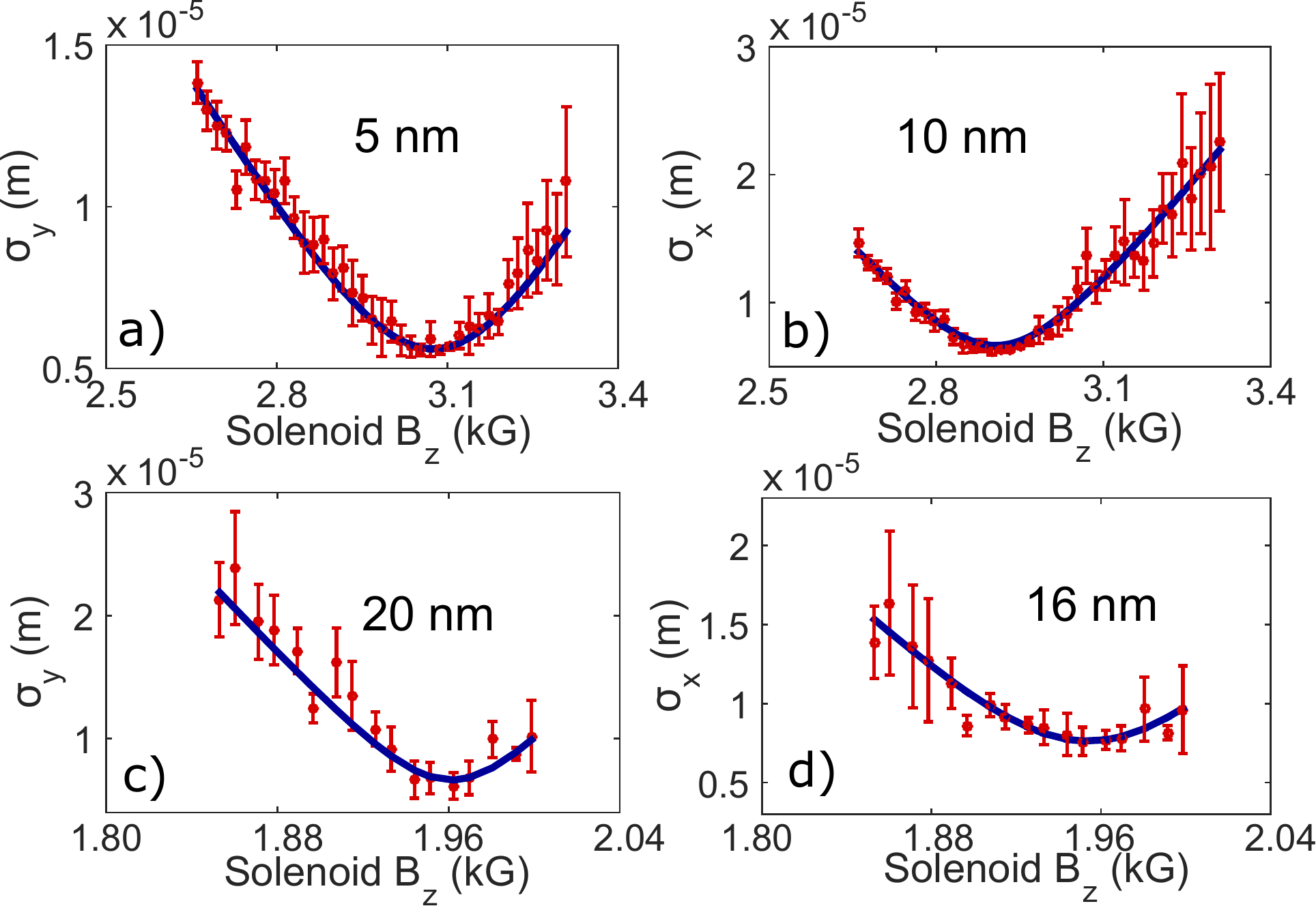}
\caption{ Normalized transverse emittances in y (a) and x (b) for a $\sigma_{t, \text{uv}} = 1.1$ ps laser pulse on the photocathode, accelerated on crest in the linac. c) and d) show normalized emittances for a $\sigma_{t, \text{uv}} = 100 $ fs, in the velocity bunching configuration. \label{emfig}}
\end{figure}

In this long laser pulse case, the emittance at 20 fC is optimized by the adjustment of the final focus lens position, as well as the field strength of the first solenoid, which is used to maintain a small spot size in the linac, limiting the emittance growth due to a spurious skew quadrupole component of the rf fields. An example solenoid scan at the optimum settings is shown in Fig. \ref{emfig} a) and b), showing transverse emittances of 5 nm $\times$ 10 nm and spot sizes down to $\sim5$ $\mu$m, indicating a very small source size and well-preserved phase space density.

LPS distributions of bunches from S-band guns typically suffer from nonlinearities induced by the sinusoidally varying field as a function of time and position. This effect is significant for velocity bunching with ps photocathode laser pulse durations, as the rf curvature can limit the minimum bunch length achievable. Promising mitigation schemes have been proposed, such as the use of a harmonic phase space linearizing cavity \cite{Floettmann2014, LiMusumeci2014}, or the use of a debunching phase in the gun to increase the spatial wavelength of the nonlinear distortions, allowing them to be linearized in the buncher \cite{ZeitlerFloettmannGruener2015}.

\begin{figure}
\includegraphics[width=0.45\textwidth]{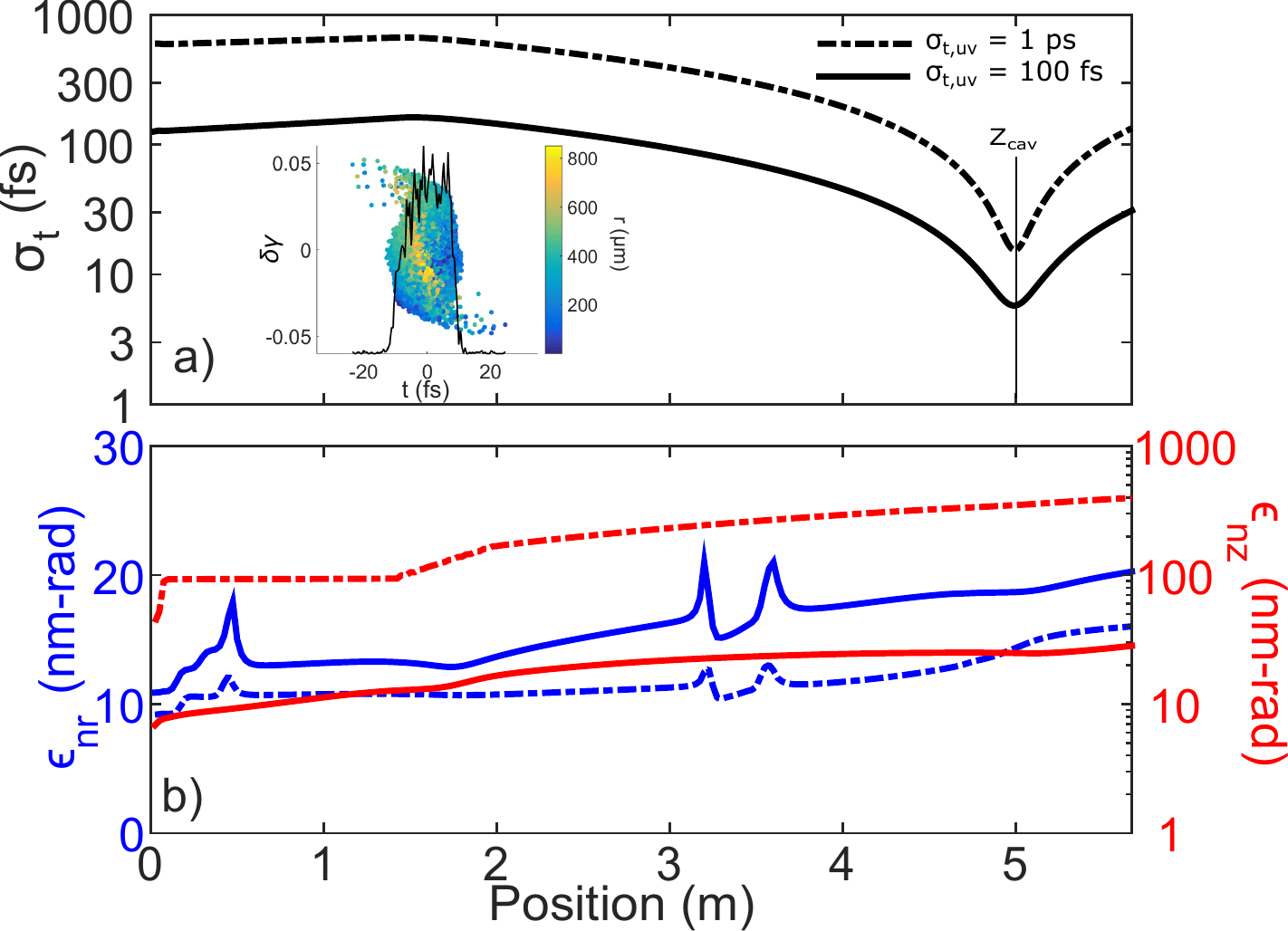}
\caption{a) Bunch length vs. position for $\sigma_{t, uv} = $ 100 fs (solid) and 1 ps (dashed). Inset: LPS and current profile of the 100 fs case at the longitudinal waist, with LPS colored by particle transverse radius. b)Evolution of longitudinal and transverse emittances along the beamline for both laser pulse lengths. \label{env} }
\end{figure}

Alternatively, one may simply employ an ultrashort laser pulse on the cathode to generate a small longitudinal emittance, given the reduced head-tail differences in the rf wave seen by the bunch. Fig. \ref{env} compares the longitudinal and transverse dynamics with a 1 ps and 100 fs rms laser pulse length, simulated with the space charge code General Particle Tracer \cite{gpt} using the transverse laser dimensions given by the virtual cathode image above. The 100 fs case approximates experimental conditions for the subsequent velocity bunching measurements at 20 fC. The longitudinal focus is placed at the deflector position, and the 1 ps case differs only by changing the focusing optics to keep the transverse and longitudinal waist positions the same. No linearization scheme is applied, hence the shorter laser pulse allows a smaller longitudinal focus (5.0 fs vs 15.5 fs) to be achieved. Note that the 100 fs case has roughly one order of magnitude smaller longitudinal emittance, comparable with the minimum emittance achieved using active compensation schemes. However, the transverse emittance in this short pulse case is diluted via increased space charge forces by a factor of 1.5 with respect to the near-intrinsic value of the long pulse case, leading to a 6D brightness $B_{6D}=Q/{\epsilon_{nx} \epsilon_{ny} \epsilon_{nz}}$ ratio between the two cases of $\sim 5$.

Motivated by this analysis, in order to demonstrate sub-10 fs bunch lengths, we reduce the laser pulse length to 100 fs rms and adjust the buncher phase to put the longitudinal focus 5 m downstream of the cathode. In this configuration, the final beam energy is reduced to 5 MeV. As in simulation, for the same beam charge, the measured transverse emittance $\sqrt{\epsilon_x \epsilon_y}$ increases. Emittance scans upstream of the longitudinal focus are shown in Fig. \ref{emfig} c) and d) yielding $\sqrt{\epsilon_x \epsilon_y} =18$ nm. The larger error bars with respect to the on-crest data are due to an observed increase in shot-to-shot energy fluctuations at the compressing phase setpoint.    

A 9 cell, 17 cm long, X-band TM110-like structure is used to measure the bunch length \cite{EnglandRosenzweigTravish2008, ScobyLiThrelkeldEtAl2013}. Deflecting cavities were previously considered to be limited in their measurement of velocity compressed beams in cases where the bunch length would vary significantly within the cavity \cite{LuTangLiEtAl2015}. However, at few MeV beam energies and when the transverse dimensions are much larger than the longitudinal dimensions, the longitudinal focus even with the inclusion of space charge forces is mostly nonlaminar \cite{Floettmann2014}. That is, as the beam comes to a longitudinal focus particles in the tail end up at the head of the beam and vice versa. This nonlaminar motion results in a cancellation of the integrated transverse momentum kicks exerted by the streaking cavity if the longitudinal focus position is placed at the center of the deflector.

The origin of this cancellation can be seen using a simple model of the streaking fields. A particle with trajectory $z = c\beta t+z_0$ will accumulate a streaked momentum
\begin{equation}
\begin{split}
\Delta p_y = \frac{F_0}{c\beta}\int_{-L/2}^{L/2}\sin \left( \frac{\omega(z-z_0)}{c \beta} \right) dz = \\
-\frac{2 F_0}{\omega} \sin \left( \frac{L\omega}{2 c \beta} \right) \sin \left( \frac{\omega z_0}{ c \beta} \right)
\end{split}
\end{equation}
where $F_0$ is the streaking force amplitude, $L$ is the cavity length, and $\omega$ is the cavity frequency. Position is measured from the center of the cavity, and the phase of the wave is chosen such that the reference particle which has $z_0 = 0$ experiences no deflection. However, any other particle with different velocity but which also has $z_0=0$ (i.e. at the cavity center at $t = 0$) also experiences no deflection. Particles with identical $z_0$ and differing velocity experience almost exactly the same kick, as the velocity spread in the beam in our case is $\delta \beta /\beta \sim 10^{-4}$. In the absence of other effects, the residual induced angular divergence is proportional to the longitudinal beam size at the center of the cavity regardless of its length at the cavity entrance and exit.

In our measurement, the beam is brought to a vertical (streak direction) focus at the final screen with a quadrupole doublet just upstream of the deflector. No slit aperture is used, contrary to other studies \cite{MusumeciMoodyEnglandEtAl2008, OudheusdenPasmansGeerEtAl2010}, and hence the projection of the entire beam distribution (and not just a slice) onto the temporal axis is measured. The vertical beam distribution is recorded with deflector on and off, $I_{on}(y)$ and $I_{off}(y)$.

The bunch length is retrieved by a detailed analysis of these traces. Assuming Gaussian beam distributions with rms width $\sigma$, the measured bunch length is then $\sigma_{t} = \frac{mc^2\beta\gamma}{\omega eV L_d} \sqrt{\sigma_{y, \text{on}}^2-\sigma_{y, \text{off}}^2}$ \cite{OudheusdenPasmansGeerEtAl2010}, where $V$ is the effective deflector voltage (400 kV), $\omega$ is the deflector angular frequency ($f = 9.6$ GHz), $L_d$ is the drift length between cavity and observation screen (68 cm), and $mc\beta\gamma$ is the average beam momentum.

\begin{figure}
\includegraphics[width=0.45\textwidth]{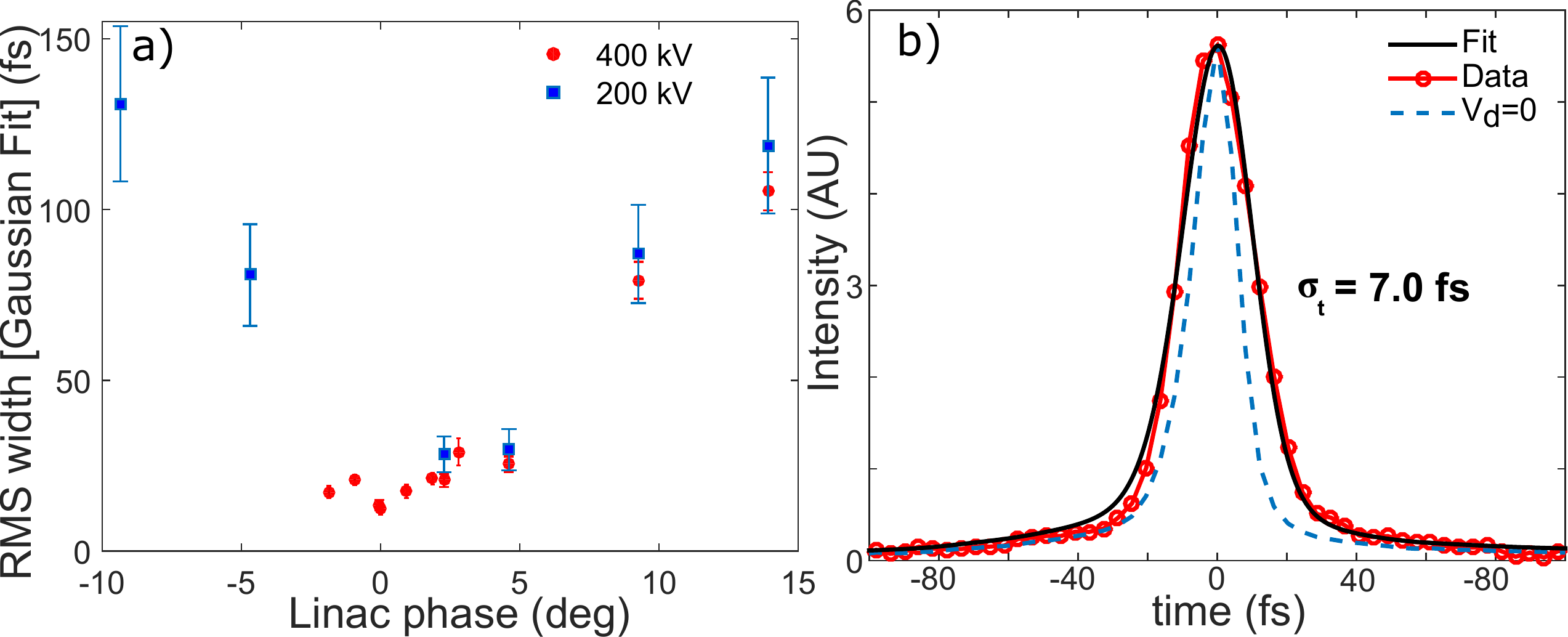}
\caption{a) Bunch length vs. relative buncher phase for deflector voltages of 400 kV and 200 kV. b) Example deflector-on shot (red), shown with mean deflector-off distribution (blue), and the result of convolution with specified $\sigma_t$ (black). \label{blens}}
\end{figure}

Using this analysis, the bunch length was measured as a function of linac phase, shown in Fig. \ref{blens}a), with statistics over roughly 10 deflector-on/off shots per phase point. The optimal bunching phase is  $\sim75^\circ$ off peak acceleration. When far from the optimal bunching phase, the deflecting voltage could be reduced to 200 kV to provide a cross-check against data taken at 400 kV.

In practice, the deflector-off distribution is not exactly Gaussian as shown in Fig. \ref{blens} where the average of all the deflector off shots shows the presence of large tails which Gaussian fits fail to capture. To get a better measure of the underlying $\sigma_t$ at the optimal buncher phase, we fit the experimental traces using $I_f(\sigma_t, t_0, t) = G(\sigma_t, t_0, t)* \langle I_{\text{off}}(t) \rangle$ to the measured $I_{\text{on}}(t)$, where $G$ is a Gaussian distribution centered at $t_0$ with rms width $\sigma_t$, $*$ is the convolution operator, $\langle I_{\text{off}}(t) \rangle$ is the average deflector off distribution. In general, the fit matches the data very well as shown in Fig. \ref{blens}, suggesting the tails in the deflector on distribution primarily arise from those present when the deflector is off. The relative variation in the width of the deflector off shots was $<5\%$ which is seen to introduce a $6\%$ rms statistical uncertainty in $\sigma_t$. The choice of a Gaussian distribution for the beam longitudinal profile was found not to be critical, as nearly identical rms bunch lengths are obtained by using a scaled simulation distribution, like the one shown in the inset of Fig. \ref{env}a. A representative shot for the optimum bunching phase is shown in Fig \ref{blens} b), yielding a 7 fs bunch length. 

It is important to consider higher order effects related to the finite transverse beam size in the measurement, given the high streaking voltage (as compared to the beam energy) and short bunch length. As a consequence of the Panofsky-Wenzel theorem \cite{Panofsky}, any deflecting structure will induce a nonzero energy spread for a finite vertical beam size in the cavity \cite{FloettmannParamonov2014, MoodyMusumeciGutierrezEtAl2009}. This induced energy spread can in turn induce bunch lengthening and defocusing, which will affect the measurement. Furthermore, transverse field nonlinearities may contribute to a position dependent streaking force and alter the beam focusing. To quantify these effects, we perform a synthetic measurement, fully analogous to that performed in experiment, using the simulation cases in Fig. \ref{env}, which have comparable transverse sizes in the deflector, along with a 3D field model of the deflector. In the long laser pulse case, with $\sigma_{t, uv} =1$ ps, the bunch length at the center of the cavity with deflector off is 15.5 fs, and the simulated measurement produces $\sigma_{t} = 15.6$ fs. Given that the $z$-average of the bunch length along the cavity is $\sim18$ fs, this illustrates that transverse effects are small here, and that the minimum bunch length can be measured in practice.

For shorter bunches, these higher order effects play a larger role. In the simulation of the $\sigma_{t, uv} = 100$ fs case, the bunch reaches a longitudinal minimum of 5.0 fs at the center of the deflector when it is off. With the cavity on, the induced energy spread causes the bunch length to grow to 6.0 fs at the center of the cavity, corresponding to $\sim 3$ fs added in quadrature, in agreement with analytical predictions \cite{FloettmannParamonov2014}. The synthetic measurement produces $\sigma_{t} = 6.7$ fs, which is additionally inflated by both transverse field nonlinearities and rf induced defocusing, and is in close agreement with the measurement in Fig \ref{blens}. Note that any focusing imparted by the deflector can lead to a systematic error in the PSF of the measurement which is otherwise given by $I_{off}(t)$. A conservative estimate of this error in the measured $\sigma_t$ for our case is $\pm 1 $ fs. This analysis shows that our measurements approach the resolution limits of the scheme for this emittance, and that reduction of the transverse emittance is critical for deflector bunch length measurements at even smaller temporal scales. Alternatively, a vertical aperture can be used to limit these effects by controlling the transverse beam dimensions into the deflector. However, this method suffers from reduced charge on the final screen, and only measures a single vertical slice of the beam.

Even though the minimum bunch length obtained is well below 10 fs, the electron bunch time-of-arrival jitter can be $>$ 30 fs, mostly from beam energy fluctuations. To apply this velocity compression scheme to ultrafast electron diffraction experiments would require a high resolution time stamping technique to temporally sort the data. Recent techniques for x-ray probes have demonstrated time-of-arrival measurements at the sub-fs level \cite{HartmannHelmlGallerEtAl2014}, and similar techniques may be applied for low charge electron probes \cite{CesarMusumeciAlesini2015}.

In conclusion, we have demonstrated generation, compression and characterization of low charge, few MeV electron beams with ($<20$ nm) emittance, and sub-10 fs bunch lengths. We describe the ability of deflecting cavities to resolve the minimum longitudinal beam size due to a transverse kick cancellation effect when the longitudinal focus is nonlaminar. This effect provides a solution to the long standing issue of how to directly measure low energy ultrashort electron beams, which will become increasingly useful with the advent of LPS linearization schemes. The transverse and longitudinal beam quality demonstrated here open the door to visualize novel ultrafast structural changes in matter, and to efficiently couple to the small acceptance of advanced accelerators.

We acknowledge G. Andonian and R. K. Li for stimulating discussions. This research was funded by NSF grant PHY-1415583 and DOE grant No. DE-SC0009914.

\end{document}